\documentclass[a4paper]{article}

\usepackage{INTERSPEECH2019}
\usepackage{booktabs}

\title{Improving multi-speaker TTS prosody variance with a residual encoder and normalizing flows}
\name{Iván Vallés-Pérez, Julian Roth, Grzegorz Beringer, Roberto Barra-Chicote, Jasha Droppo}
\address{
  Alexa AI}
\email{\{ivallesp,julroth,beringg,rchicote,drojasha\}@amazon.com}

\begin{document}

\maketitle
\begin{abstract}
Text-to-speech systems recently achieved almost indistinguishable quality from human speech. However, the prosody of those systems is generally flatter than natural speech, producing samples with low expressiveness. Disentanglement of speaker id and prosody is crucial in text-to-speech systems to improve on naturalness and produce more variable syntheses. This paper proposes a new neural text-to-speech model that approaches the disentanglement problem by conditioning a \textit{Tacotron2}-like architecture on flow-normalized speaker embeddings, and by substituting the reference encoder with a new learned latent distribution responsible for modeling the intra-sentence variability due to the prosody. By removing the reference encoder dependency, the speaker-leakage problem typically happening in this kind of systems disappears, producing more distinctive syntheses at inference time. The new model achieves significantly higher prosody variance than the baseline in a set of quantitative prosody features, as well as higher speaker distinctiveness, without decreasing the speaker intelligibility. Finally, we observe that the normalized speaker embeddings enable much richer speaker interpolations, substantially improving the distinctiveness of the new interpolated speakers.
\end{abstract}

\vspace{-0.1cm}
\section{Introduction}
In the last five years speech technologies have improved considerably. The text-to-speech (TTS hereafter) field has been largely benefited by the rise of deep learning \cite{Sisman2021}, which allowed these systems to achieve near-human performance at synthesizing speech that is almost indistinguishable from human's. One of the first achievements was \textit{WaveNet} \cite{vanderoord2016}, a neural vocoder based on dilated causal convolutions that was able to surpass its predecessors in naturalness in 2016. With the rise of the attention mechanism and its variants \cite{bahdanau2015,vaswani2017,chaudhari2019}, \textit{Tacotron} \cite{Wang2017} and \textit{Tacotron 2} \cite{Shen2018,Liu2019} proposed a sequence to sequence architecture as an end-to-end TTS solution. These models were able to map input text (or phonemes) to a spectrogram that, given the right neural vocoder (\textit{WaveNet} for instance), would be converted into a sequence of waveform samples. 

TTS is fundamentally a generative modeling problem because a given sentence can be mapped to multiple utterances with different prosody and speaker characteristics \cite{Taylor2009}. \textit{Tacotron}, in its initial form, is a supervised model that performs a hard mapping between the input text and the output spectrograms. Therefore, the utterances generated using \textit{Tacotron}-like architectures, although natural, they follow the average prosody of the training set, not allowing to generate utterances of a sentence with multiple speaking styles.   

The authors of \cite{skerryryan2018} attempt to remediate the lack of expressiveness of the model by introducing a reference encoder consisting of a latent distribution that is conditioned on a reference mel-spectrogram (usually the target spectrogram, at training time). This distribution is learned by the model through a bottleneck that is intended to capture prosody aspects of the target spectrogram, preventing phonetic or speaker information to flow through. In practice, specially when using this approach in multi-speaker settings, the model tends to leak speaker information to the output. This represents a problem at inference time, when a synthetic neutral reference is provided (generally synthesized using a production system similar to \textit{Amazon Polly}), given that the synthesized speaker identity tends to resemble the reference instead of the target voice. This problem is known as speaker leakage and it was already reported in \cite{skerryryan2018}, where the authors emphasize the importance of properly tuning the size of the reference bottleneck to amend it.

In \cite{Liu2020}, the authors propose using a multi-task version of \textit{Tacotron} to enhance the prosody of the syntheses. The approach described in the paper consists of jointly learning the target mel-spectrogram as well as the probability distribution of the phrase break patterns for each word. The work of \cite{liu2020expressive} proposes a novel training schema for \textit{Tacotron} where deep style features are extracted using the \textit{SER} framework (\cite{Zhang2018, Lotfian2019}). These features are later used for minimizing the style differences between the real and synthesized samples.  

In this work, we start from a multi-speaker \textit{Tacotron} architecture with a reference encoder (similar to \cite{skerryryan2018}) and propose two modifications: (1) we replace the pre-trained speaker embedding with a normalized speaker embedding using normalizing flows \cite{Kingma2018}, that allows sampling from the learned \textit{Gaussian} distribution and (2) we substitute the reference encoder with a residual encoder, which learns a latent distribution conditioned on the input phonetic information, allowing the model to capture and encode the residual attributes not present in the linguistic input and the speaker embedding. 

Inspired by the work of \cite{Raitio2020}, we show how our proposed model achieves significantly higher prosody variation by measuring the difference in variance between the baseline and the proposed model across a set of features derived from the fundamental frequency, the energy, the signal to noise ratio and the speaking rate. These features capture most of the prosody variables (pitch, speed, loudness and timbre) \cite{Raitio2020} and allow performing objective comparisons between systems.

Our contribution can be summarized in the following points: (1) we propose a new architecture that significantly increases the prosody variability of the syntheses with no intelligibility and distinctiveness degradation, (2) the proposed architecture removes the dependency on a production voice system as it does not use a reference spectrogram, (3) when interpolating between speaker embeddings the synthetic speakers from the proposed model are significantly more distinct than the ones of the baseline.

\section{Methods}
\begin{figure}
	\centering
	
	\includegraphics[width=1.0\linewidth]{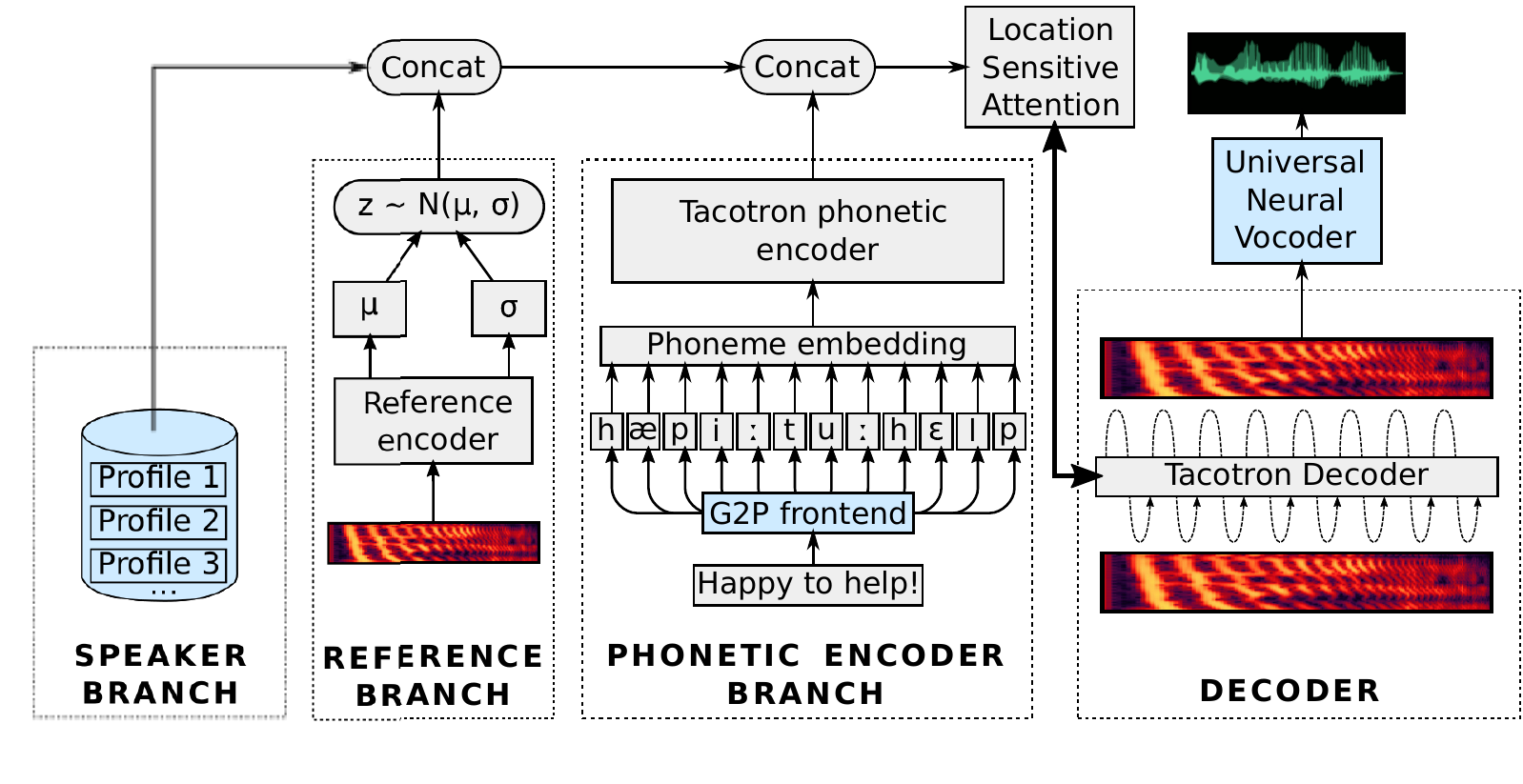}
	
	\vspace{0.5cm}
	
	\includegraphics[width=1.0\linewidth]{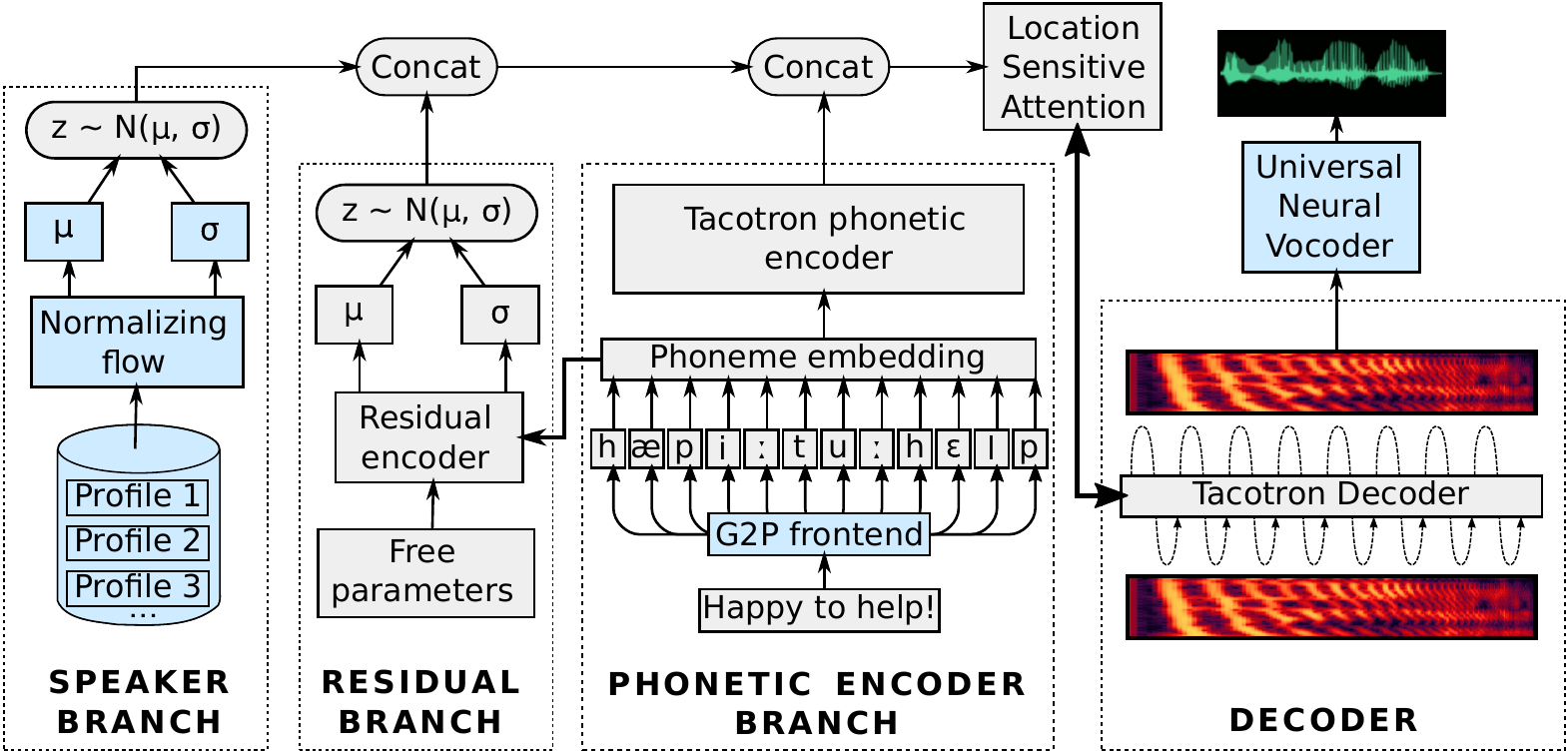}
	
	\caption{Top: the baseline architecture. Bottom: the proposed architecture. The blue blocks represent pretrained parts of the network and the blocks with round edges are tensor operations.}
	\label{fig:architectures}
\end{figure}

A \textit{Tacotron2}-based sequence to sequence model with \textit{location-based attention} \cite{luong2015effective} is used as a baseline throughout this study. This baseline architecture is represented in figure \ref{fig:architectures}-top. A couple of encoder branches are included over the initial \textit{Tacotron} formulation to allow the model to synthesize multiple speakers: the speaker branch and the reference branch. The first one takes as input a pre-trained speaker embedding vector representing the speaker characteristics. This vector is obtained from the output of a speaker verification model trained to minimize a triplet loss. This speaker verification model is pre-trained using pairs of utterances, similar to  \cite{Ren2019}. For every speaker, the vectors corresponding to all their utterances are pre-computed and then averaged to form the speaker vectors. As a result, we have a fixed size vector for every speaker in the data set. Additionally, we include a reference branch that is conditioned on the target spectrogram (at training time) and is intended to learn a latent distribution summarizing the prosody information, similar to the proposal of \cite{skerryryan2018}.

% TODO: \usepackage{graphicx} required

The proposed architecture builds upon the baseline model. We remove the dependency of the reference branch on the target spectrogram, because it tends to cause speaker/phonetic leakage (read the section 4 of the samples included in \cite{skerryryan2018}). It also slows down the inference process as it requires a production TTS system to provide the references. Instead, we introduce a learnable variational latent space conditioned on the phonetic information and a set of learnable free parameters that we name \textbf{residual branch}. The purpose of this branch is to encode the prosody information not present in the input linguistic features or in the speaker embedding vector in a new latent space (i.e. the residual prosody variance). Moreover, we normalize the speaker embedding vectors using normalizing flows based on the work of \cite{Kingma2018}, so that the normalized vectors follow a \textit{Gaussian} distribution.That change enables sampling from the speaker embedding latent space instead of just using the average embedding vector. This increases the coverage of the speaker embedding space, allowing smoother interpolations between speaker embedding vectors. The architecture proposed is depicted in figure \ref{fig:architectures}-bottom

\subsection{Speaker embedding normalization using normalizing flows}
Normalizing flows \cite{rezende2016variational} are powerful methods for producing tractable distributions from which one can sample. Through a sequence of invertible transformations they transform a complex input distribution to a tractable probability distribution. The output distribution is usually chosen to be an isotropic unit \textit{Gaussian} to allow for smooth interpolations and efficient sampling. 

\textit{Glow} \cite{Kingma2018}, a flow-based generative model, has shown significant improvements in computer vision generative modeling by adding invertible 1x1 convolutions to the sequence of transformations applied to the input. To train this model on one-dimensional speaker embedding vectors, we have switched the 2-dimensional convolutions with 1-dimensional ones.

We train a normalizing flow over the pre-trained speaker embeddings described previously. The trained model attains a \textit{Gaussian} distributed latent space of speaker embeddings from which one can easily sample to create embeddings representing new, unseen speakers. The proposed architecture samples from the \textit{Gaussian} distribution defined by the normalized embeddings, both at training time and at inference time.

\subsection{Residual branch}
The new residual branch is designed to learn the prosody-induced variance that cannot be explained by the linguistic features and the speaker embedding vector alone. It takes as input the linguistic features and a set of learnable free parameters\footnote{Notice that the dense layers that produce the residual latent distribution parameters do not have bias terms, so that all the bias is learned in the free parameters vector.}, conditioning on the target sentence and on a global prior. The motivation behind this design is to help the network learn different ways of uttering a given input sentence (linguistic features conditioning), and the potential global biases existing in our training dataset (free parameters).

The output of the phonetic encoder is piped into a bi-directional recurrent neural network \cite{Schuster1997,graves2005} in order to get a representation independent of time. The last output of the recurrent neural network of both, forward and reverse passes (represented in equations \ref{eq:rnn1} and \ref{eq:rnn2} as $\mathbf{o_f}$ and $\mathbf{o_r}$, respectively), are  concatenated to form a fixed-size phonetic representation $[\mathbf{o_f}, \mathbf{o_r}]$. That vector is concatenated with a vector of free parameters $\mathbf{v_f}$ and the result is passed through two stacked dense layers to form the parameters of the residual latent distribution (as shown in equation \ref{eq:residual}, where $f$ represents the \textit{ReLU} activation function). The $h_{residual}$ vector is split in two vectors $h_{residual}^{\mu}$ and $h_{residual}^{\sigma}$ from where a latent vector $z$ is sampled using the re-parametrization trick \cite{Kingma2019}. We add a \textit{Kullback-Leibler} loss to assure that the distribution of the latent representation approximates a \textit{Gaussian} distribution. 

\begin{equation}
 \label{eq:rnn1}
\mathbf{o_f} = \text{RNN}_f(\mathbf{h_{ph}})
\end{equation}

\begin{equation}
 \label{eq:rnn2}
\mathbf{o_r} = \text{RNN}_r(\mathbf{h_{ph}})
\end{equation}

\begin{equation}
 \label{eq:residual}
\mathbf{h_{residual}} = \mathbf{W_2}(f(\mathbf{W_1}\cdot([\mathbf{o_f}, \mathbf{o_r}, \mathbf{v_f}]) + \mathbf{b1} )) +\mathbf{ b_2}
\end{equation}

\section{Experiments}
 We have used a combination of two internal datasets to train the models, one of them containing 2860 non-professional speakers, with 200 utterances per speaker on average, and the other containing 10 professional speakers, with 13,000 studio-recorded utterances per speaker – totaling to 2870 speakers and more than 700,000 utterances. We process audio utterances at 16kHz and extract 80 dimensional mel-spectrograms. We define a frame as a $50ms$ sequence, with an overlap of $12.5ms$. We used the Universal Neural vocoder to synthesize the wav samples \cite{lorenzotrueba2019}. The linguistic features have been extracted using an internal linguistic front-end which takes the text as input and extracts the phonemes, that are used as input for the model. The speaker embedding, the free parameters and the residual distribution parameters $\mu$ and $\sigma$ vectors were defined to have a length of 192 elements.
 
 We have trained both models on one million batches, with a batch size of 24. Then we synthesized a set of 50 unseen sentences for each of the 2870 speakers. The syntheses have been evaluated in 3 ways: intelligibility, distinctiveness and prosody variability.
 
 Additionally, we sampled 50 speakers randomly from the pool of 2870 speakers and generated interpolations between all the possible pairs, in order to study how the proposed model and the baseline behave in this setting.
 
 \subsection{Intelligibility}
We used the \textit{AWS transcribe} system to transcribe each of the syntheses, and then measure the Word Error Rate (\textit{WER}) between the target sentence and the transcription \cite{Uday2019}. These metrics are aggregated as the average \textit{WER} per speaker. We achieved a median Word Error Rate of $8.5\%$ in the baseline, and $8.3\%$ in the proposed architecture. Although significative, this difference is very small, thus we can conclude that both models are roughly equivalent in terms of intelligibility.

% TODO: \usepackage{graphicx} required
\begin{figure}[h]
	\centering
	\includegraphics[width=0.95\linewidth]{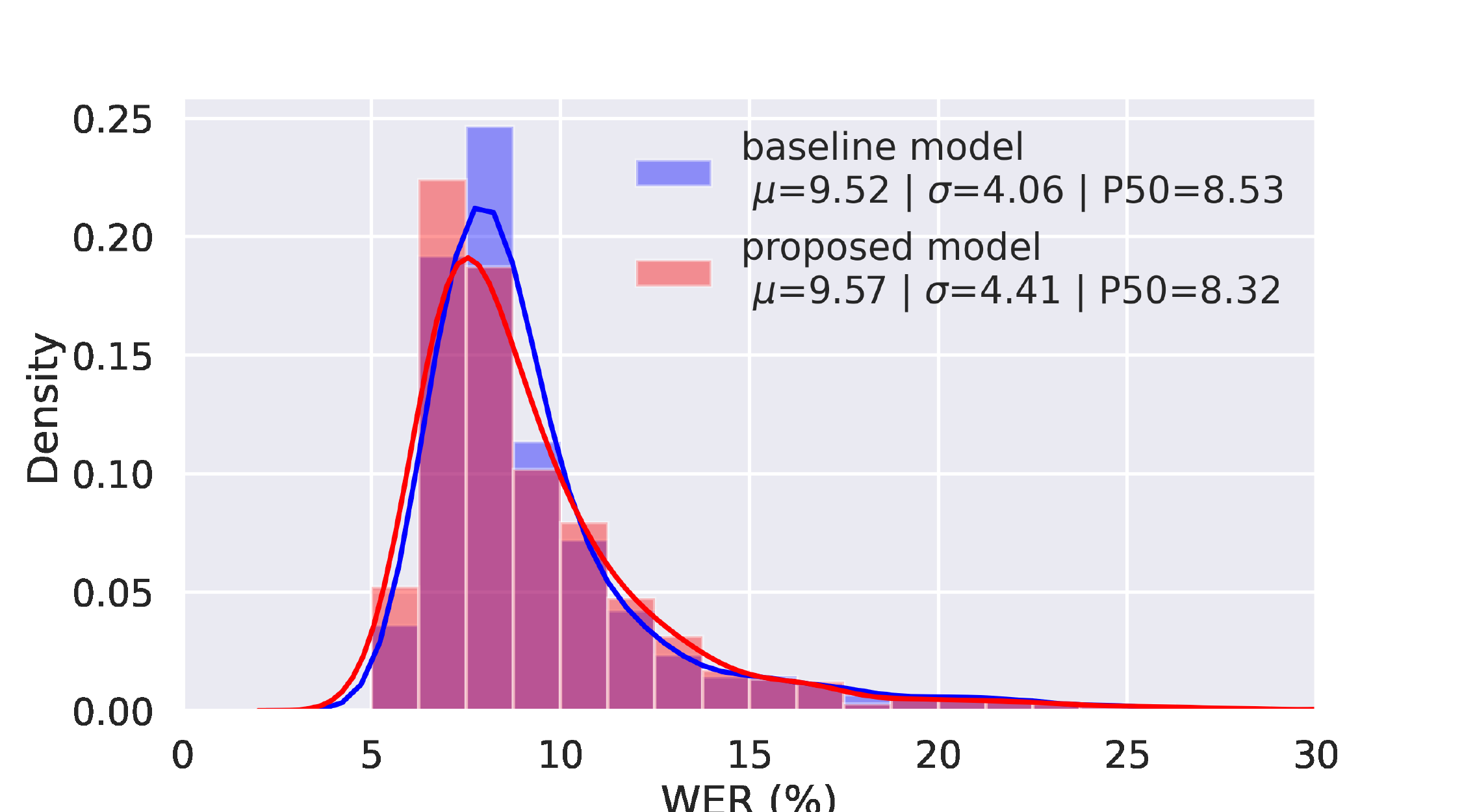}
	\caption{Speaker intelligibility of the baseline and residual models, measured as Word Error Rate of the target and the transcribed synthesis, showing a slightly smaller error for the proposed architecture (Wilcoxon's $p=0.0305$).}
	\label{fig:wer}
\end{figure}

\vspace{-0.3cm}
\subsection{Distinctiveness}
\begin{figure}[h]
	\centering
	\includegraphics[width=.95\linewidth]{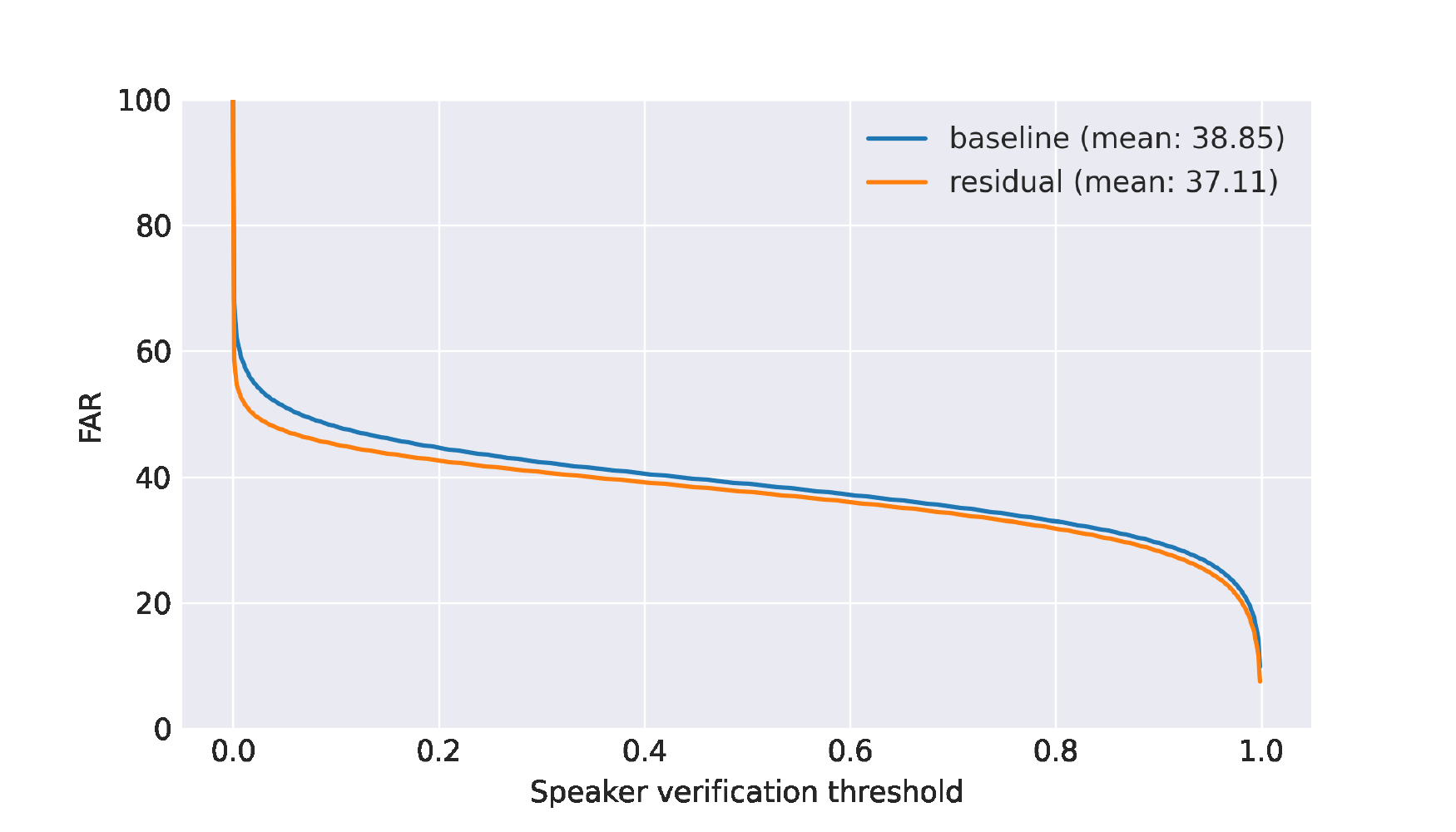}
	\caption{False Acceptance Rate distribution at different \textit{speaker verification model} thresholds.}
	\label{fig:far}
\end{figure}

To measure how distinct the synthesized speakers are, we use our internal \textit{speaker verification system}.  Each utterance is compared with 4 samples randomly drawn from the full pool of samples. We measure the False Acceptance Rate (\textit{FAR}), which quantifies the percentage of speaker pairs incorrectly identified as the same speakers by the speaker verification model. Figure \ref{fig:far} shows how the \textit{FAR} metric varies for both, the proposed model and the baseline as we vary the classification threshold. Table \ref{tab:far} shows the \textit{FAR} score for four arbitrarily picked thresholds. Lower values of \textit{FAR} mean higher distinctiveness.

\begin{table}[h]
	\centering
	\caption{False Acceptance Rate metric for different thresholds}
	\footnotesize
	\begin{tabular}{ccc}
		\toprule
		\textbf{Threshold} & \textbf{FAR-Baseline model} & \textbf{FAR-Proposed model} \\
		\midrule
		85\% & 31.49\% & 30.27\% \\
		
		90\% & 29.50\% & 28.21\% \\
		
		95\% & 26.23\% & 24.77\% \\
		
		99\% & 19.05\% & 16.93\% \\
		\bottomrule
	\end{tabular}
	\label{tab:far}
\end{table}

\vspace{-0.3cm}

\subsection{Prosody variability}
To measure how much prosody and quality variance the new model introduces with respect to the baseline, we have defined a set of objective metrics inspired on the work of \cite{Raitio2020}. Those metrics are listed in table \ref{tab:metrics}.

% TODO: \usepackage{graphicx} required
\begin{figure}[h!]
	\centering
	\includegraphics[width=0.7\linewidth]{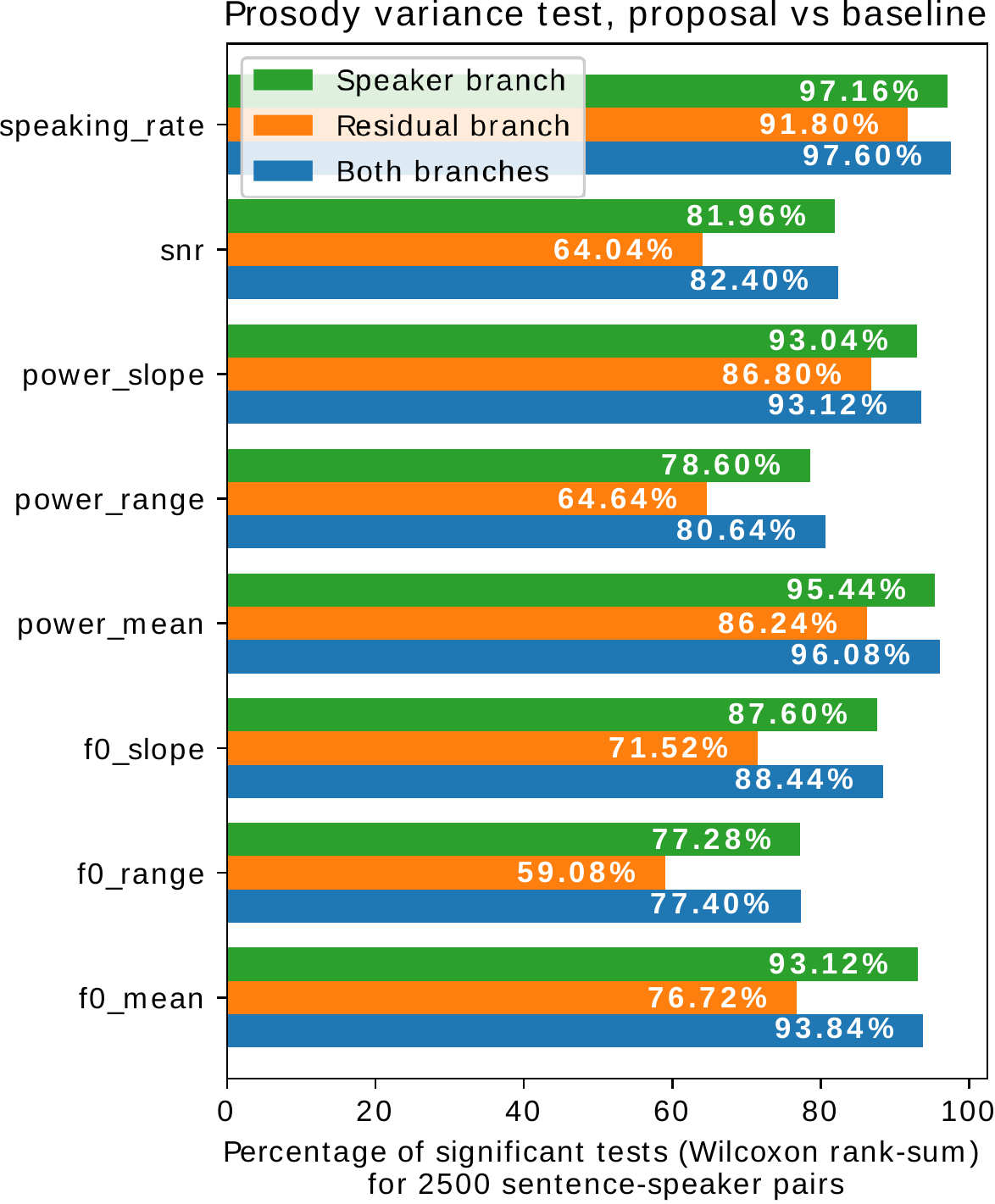}
	\caption{Baseline model vs proposed model when sampling from the residual latent distribution (blue) or from the speaker embedding distribution (orange), keeping the other constant. The results are represented as proportion of speaker-sentence pairs for which the proposed model shows significantly higher variance than the baseline, for each of the previously defined prosody features.}
	\label{fig:wilcoxon}
\end{figure}

\begin{table*}
	\footnotesize
		\centering
	\caption{Metrics used to quantify the prosody variation and quality across different models.}
	\begin{tabular}{p{0.1\textwidth}|p{0.8\textwidth}}
		\toprule
		Metric & Description \\
		\midrule
		f0 mean &  Mean of the f0 (calculated using SPTK \cite{sptk} considering only the vocal sounds (non-zero components), with frame skip of 12.5ms.\\
		f0 range & Difference between the 5th and 95th percentiles of f0. \\
		speaking rate & Calculated as the number of phonemes in the sentence divided by total duration of the synthesis. \\
		f0 slope & Calculated as the slope of a linear fit using least squares in the f0 plot. \\
		snr & Signal to noise ratio calculated using SOX \cite{SOX} as the difference in RMS dBs between the loudest and the quietest windows, using windows size of 50ms. \\
		power mean & $20\log10(\hat{x})$, where $\hat{x}$ is the average absolute amplitude with frame skip of 12.5ms. Only considering the vocal sounds. \\
		power range & Difference between the 5th and 95th percentiles of the power along time. \\
		power slope & Calculated as the slope of a linear fit using least squares in the power plot. \\
		\bottomrule
	\end{tabular}
	\label{tab:metrics}
\end{table*}
	
For this experiment, we have picked 50 speakers and 50 sentences. Then, 30 samples were synthesized for every speaker-sentence pair varying the random seed, so different latent vectors are drawn from the latent distributions in each repetition. We repeat this procedure four times: (1) with the baseline model, (2) with the proposed model multiplying the $\sigma$ parameter of the speaker embedding distribution by zero (so that the latent vector becomes the mean of the speaker embedding latent distribution), (3) with the proposed model multiplying the $\sigma$ parameter of the residual distribution by zero, and (4) allowing sampling from both branches in the proposed model. We then compare the syntheses of (1) vs (2), (1) vs (3) and (1) vs (4) .

For the comparison, as we are interested to measure differences of variance for the variables defined in the table, we first use \textit{bootstrap} to approximate the variance distribution for every variable in every speaker-sentence pair, and then we run a one-way \textit{Wilcoxon Signed-Rank} sum test between the groups. We use a significance level of $\alpha=5\%$ to which we apply the \textit{Bonferroni} correction ($\alpha=0.05/2500=0.002\%$). The results of the tests are summarized in the figure \ref{fig:wilcoxon}.

We have informally observed by listening to the syntheses that when we sample from the residual distribution (keeping the speaker embedding distribution constant), the variations in the syntheses are related with prosody aspects like the speaking rate, syllable duration or intonation, keeping, the speaker identity is untouched. When sampling from the speaker embedding distribution, we notice small variations on the speaker identity. 

\vspace{-0.1cm}

\subsection{Speakers interpolation}
\begin{figure}[h]
	\centering
	\includegraphics[width=0.94\linewidth]{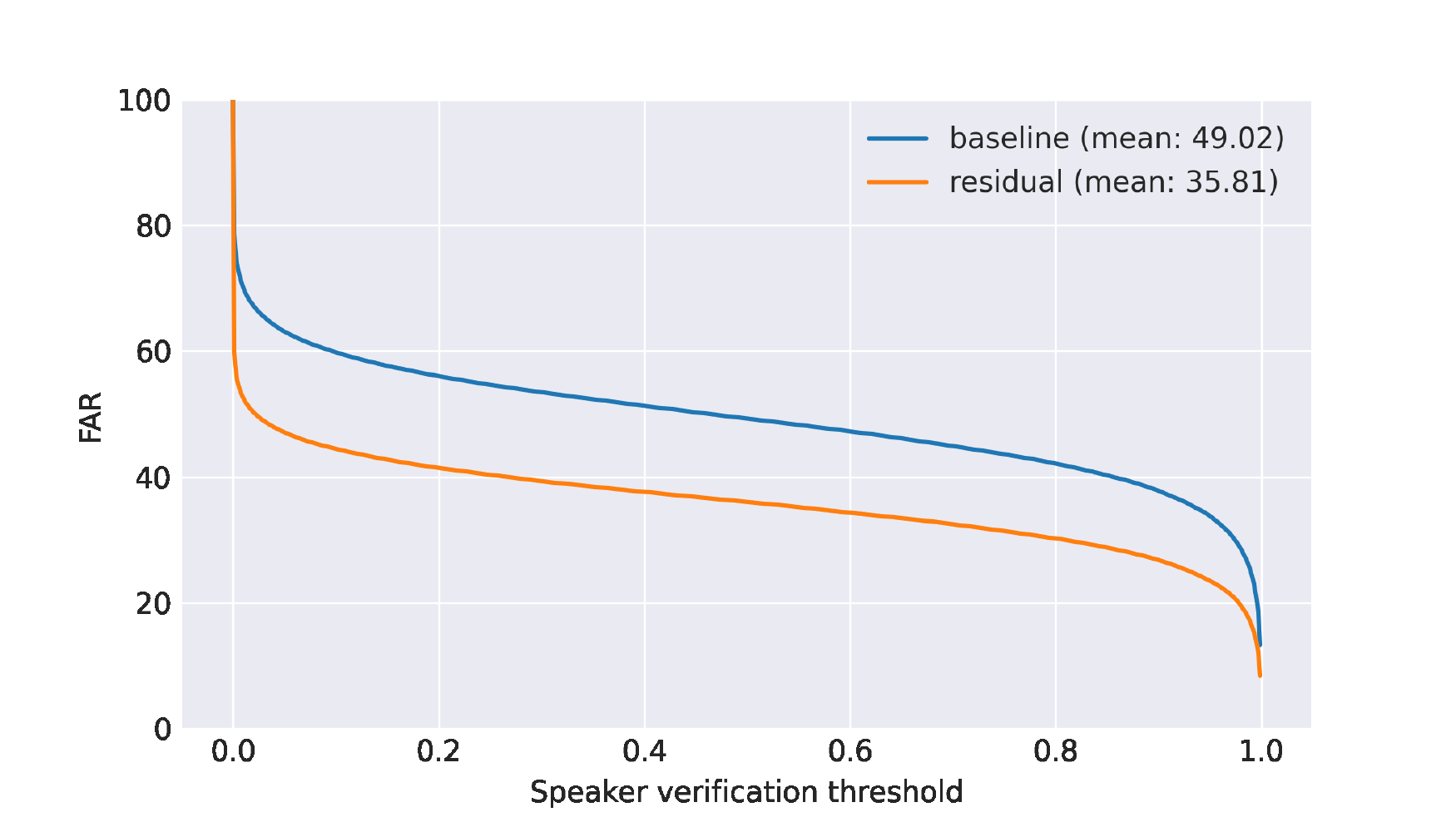}

	\includegraphics[width=0.94\linewidth]{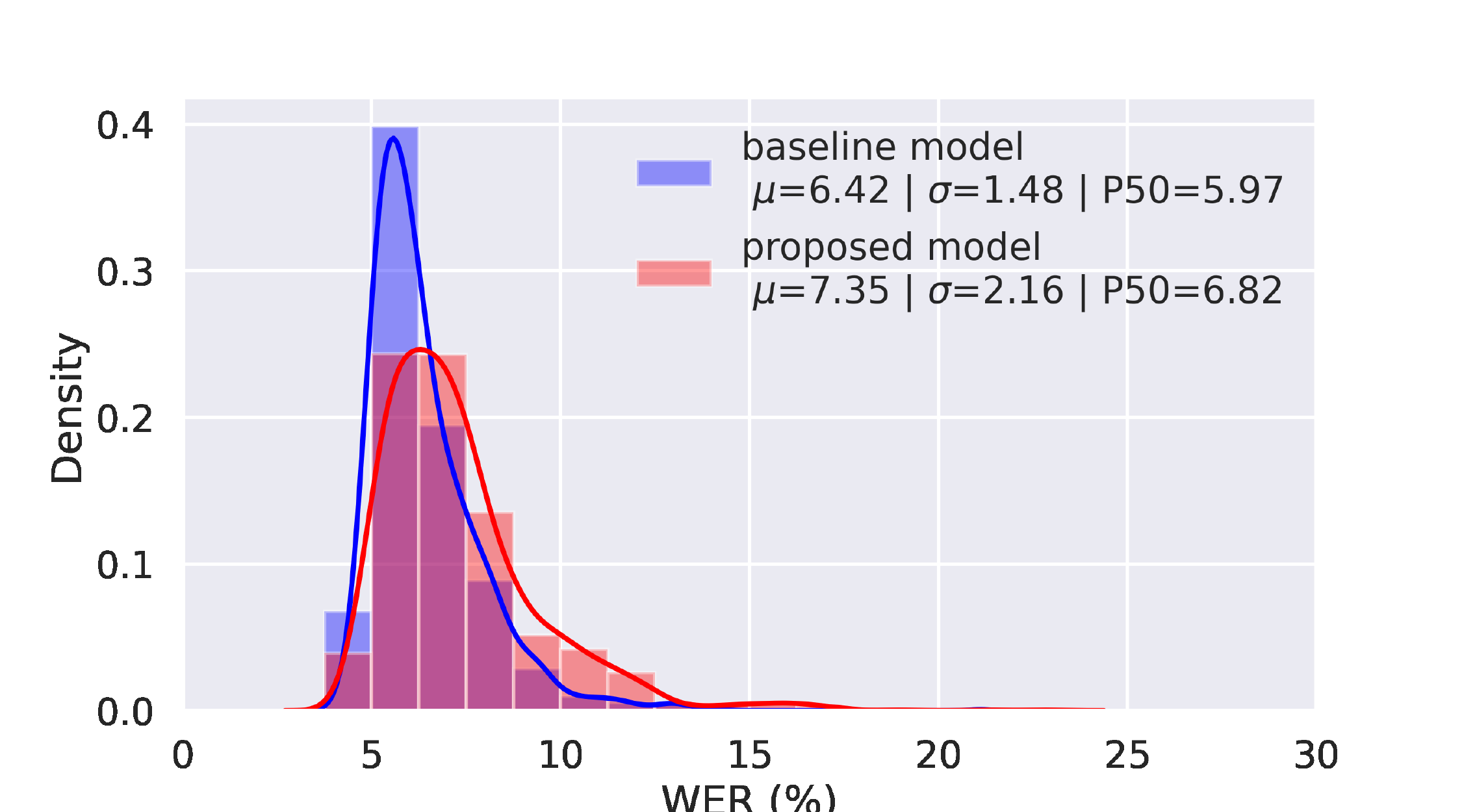}
	\caption{Top: False Acceptance Rate distribution at different speaker verification thresholds, for all the interpolated speakers. Bottom: speaker intelligibility of the baseline and residual models, measured as Word Error Rate of the target and the transcribed synthesis, showing a slightly smaller error for the baseline architecture (Wilcoxon's $p=5.2263\cdot 10^{-6}$)}
	\label{fig:svps}
\end{figure}

We have tested the proposed model and the baseline using interpolated speakers to study how the speaker embedding normalization affects the intelligibility and distinctiveness of the new speakers. For that, 50 speakers have been chosen. We have interpolated all the pairs of speakers using polar interpolation, generating 1225 new voice profiles. We have synthesized 50 sentences using those new voices and then evaluated distinctiveness and intelligibility (see results in figure  \ref{fig:svps}).
% TODO: \usepackage{graphicx} required

The proposed model achieves 13.11\% lower FAR than the baseline. We hypothesize that the rationale behind this is that the normalized speaker embedding space has a denser latent distribution than the one in the baseline. Given that the proposed model is trained on samples drawn from the normalized speaker embedding distribution (as opposed of using average vectors as in the baseline model), the latent space has higher density and hence the generalization to new speakers is better.

From the intelligibility perspective, the baseline shows an average WER of $6.42\%$ while the proposed model achieved a $7.35\%$. We attribute that difference to the fact that the interpolated speakers in the baseline are less distinctive and resemble much more to one of the two actual speakers, hence its intelligibility is naturally higher at the cost of a worse distinctiveness. Although significative, the difference in practice is negligible.

\vspace{-0.3cm}
\section{Conclusions}
\vspace{-0.1cm}

We introduced a new TTS architecture that allows increasing the prosody variance of a multi-speaker TTS system by learning the residual prosody into a new latent distribution. We showed that by sampling from the residual and normalized speaker latent distribution the model produces syntheses with significantly different prosodies as measured by a set of quantitative metrics. The inclusion of the residual distribution also enables removing the reference spectrogram dependency. This does not only solve potential speaker leakage issues but also allows for a faster and potentially cheaper inference, given that the model does not depend on a voice production system to work.

Finally, the normalization of the speaker embedding latent space allows for better speaker interpolation when compared with the baseline model, thus producing more diverse and unseen synthetic speakers.

\bibliographystyle{IEEEtran}

\bibliography{mybib}

\begin{thebibliography}{10}
\providecommand{\url}[1]{#1}
\csname url@samestyle\endcsname
\providecommand{\newblock}{\relax}
\providecommand{\bibinfo}[2]{#2}
\providecommand{\BIBentrySTDinterwordspacing}{\spaceskip=0pt\relax}
\providecommand{\BIBentryALTinterwordstretchfactor}{4}
\providecommand{\BIBentryALTinterwordspacing}{\spaceskip=\fontdimen2\font plus
\BIBentryALTinterwordstretchfactor\fontdimen3\font minus
  \fontdimen4\font\relax}
\providecommand{\BIBforeignlanguage}[2]{{%
\expandafter\ifx\csname l@#1\endcsname\relax
\typeout{** WARNING: IEEEtran.bst: No hyphenation pattern has been}%
\typeout{** loaded for the language `#1'. Using the pattern for}%
\typeout{** the default language instead.}%
\else
\language=\csname l@#1\endcsname
\fi
#2}}
\providecommand{\BIBdecl}{\relax}
\BIBdecl

\bibitem{Sisman2021}
B.~{Sisman}, J.~{Yamagishi}, S.~{King}, and H.~{Li}, ``An overview of voice
  conversion and its challenges: From statistical modeling to deep learning,''
  \emph{IEEE/ACM Transactions on Audio, Speech, and Language Processing},
  vol.~29, pp. 132--157, 2021.

\bibitem{vanderoord2016}
\BIBentryALTinterwordspacing
A.~van~den Oord, S.~Dieleman, H.~Zen, K.~Simonyan, O.~Vinyals, A.~Graves,
  N.~Kalchbrenner, A.~Senior, and K.~Kavukcuoglu, ``Wavenet: A generative model
  for raw audio,'' in \emph{Arxiv}, 2016. [Online]. Available:
  \url{https://arxiv.org/abs/1609.03499}
\BIBentrySTDinterwordspacing

\bibitem{bahdanau2015}
D.~Bahdanau, K.~Cho, and Y.~Bengio, ``Neural machine translation by jointly
  learning to align and translate,'' in \emph{3rd International Conference on
  Learning Representations}, ser. ICLR'15, July 2015.

\bibitem{vaswani2017}
A.~Vaswani, N.~Shazeer, N.~Parmar, J.~Uszkoreit, L.~Jones, A.~N. Gomez,
  u.~Kaiser, and I.~Polosukhin, ``Attention is all you need,'' in
  \emph{Proceedings of the 31st International Conference on Neural Information
  Processing Systems}, ser. NIPS'17, December, 2017, p. 6000–6010.

\bibitem{chaudhari2019}
\BIBentryALTinterwordspacing
S.~Chaudhari, G.~Polatkan, R.~Ramanath, and V.~Mithal, ``An attentive survey of
  attention models,'' \emph{CoRR}, vol. abs/1904.02874, 2019. [Online].
  Available: \url{http://arxiv.org/abs/1904.02874}
\BIBentrySTDinterwordspacing

\bibitem{Wang2017}
\BIBentryALTinterwordspacing
Y.~Wang, R.~Skerry-Ryan, D.~Stanton, Y.~Wu, R.~J. Weiss, N.~Jaitly, Z.~Yang,
  Y.~Xiao, Z.~Chen, S.~Bengio, Q.~Le, Y.~Agiomyrgiannakis, R.~Clark, and R.~A.
  Saurous, ``Tacotron: Towards end-to-end speech synthesis,'' 2017. [Online].
  Available: \url{https://arxiv.org/abs/1703.10135}
\BIBentrySTDinterwordspacing

\bibitem{Shen2018}
\BIBentryALTinterwordspacing
J.~Shen, R.~Pang, R.~J. Weiss, M.~Schuster, N.~Jaitly, Z.~Yang, Z.~Chen,
  Y.~Zhang, Y.~Wang, R.~Skerry-Ryan, R.~A. Saurous, Y.~Agiomyrgiannakis, and
  Y.~Wu, ``Natural tts synthesis by conditioning wavenet on mel spectrogram
  predictions,'' 2018. [Online]. Available:
  \url{https://arxiv.org/abs/1712.05884}
\BIBentrySTDinterwordspacing

\bibitem{Liu2019}
\BIBentryALTinterwordspacing
R.~Liu, B.~Sisman, J.~Li, F.~Bao, G.~Gao, and H.~Li, ``Teacher-student training
  for robust tacotron-based {TTS},'' vol. abs/1911.02839, 2019. [Online].
  Available: \url{http://arxiv.org/abs/1911.02839}
\BIBentrySTDinterwordspacing

\bibitem{Taylor2009}
P.~Taylor, \emph{Text-to-Speech Synthesis}, 1st~ed.\hskip 1em plus 0.5em minus
  0.4em\relax USA: Cambridge University Press, 2009.

\bibitem{skerryryan2018}
\BIBentryALTinterwordspacing
R.~Skerry-Ryan, E.~Battenberg, Y.~Xiao, Y.~Wang, D.~Stanton, J.~Shor, R.~Weiss,
  R.~Clark, and R.~A. Saurous, ``Towards end-to-end prosody transfer for
  expressive speech synthesis with tacotron,'' in \emph{Proceedings of the 35th
  International Conference on Machine Learning}, ser. Proceedings of Machine
  Learning Research, J.~Dy and A.~Krause, Eds., vol.~80.\hskip 1em plus 0.5em
  minus 0.4em\relax Stockholmsmässan, Stockholm Sweden: PMLR, 10--15 Jul 2018,
  pp. 4693--4702. [Online]. Available:
  \url{http://proceedings.mlr.press/v80/skerry-ryan18a.html}
\BIBentrySTDinterwordspacing

\bibitem{Liu2020}
\BIBentryALTinterwordspacing
R.~Liu, B.~Sisman, F.~Bao, G.~Gao, and H.~Li, ``Modeling prosodic phrasing with
  multi-task learning in tacotron-based tts,'' \emph{IEEE Signal Processing
  Letters}, vol.~27, p. 1470–1474, 2020. [Online]. Available:
  \url{http://dx.doi.org/10.1109/LSP.2020.3016564}
\BIBentrySTDinterwordspacing

\bibitem{liu2020expressive}
R.~Liu, B.~Sisman, G.~Gao, and H.~Li, ``Expressive tts training with frame and
  style reconstruction loss,'' 2020.

\bibitem{Zhang2018}
S.~{Zhang}, S.~{Zhang}, T.~{Huang}, and W.~{Gao}, ``Speech emotion recognition
  using deep convolutional neural network and discriminant temporal pyramid
  matching,'' \emph{IEEE Transactions on Multimedia}, vol.~20, no.~6, pp.
  1576--1590, 2018.

\bibitem{Lotfian2019}
R.~{Lotfian} and C.~{Busso}, ``Curriculum learning for speech emotion
  recognition from crowdsourced labels,'' \emph{IEEE/ACM Transactions on Audio,
  Speech, and Language Processing}, vol.~27, no.~4, pp. 815--826, 2019.

\bibitem{Kingma2018}
D.~P. Kingma and P.~Dhariwal, ``Glow: Generative flow with invertible 1x1
  convolutions,'' in \emph{Advances in Neural Information Processing Systems},
  S.~Bengio, H.~Wallach, H.~Larochelle, K.~Grauman, N.~Cesa-Bianchi, and
  R.~Garnett, Eds., vol.~31.\hskip 1em plus 0.5em minus 0.4em\relax Curran
  Associates, Inc., 2018, pp. 10\,215--10\,224.

\bibitem{Raitio2020}
\BIBentryALTinterwordspacing
T.~Raitio, R.~Rasipuram, and D.~Castellani, ``Controllable neural
  text-to-speech synthesis using intuitive prosodic features,'' 2020. [Online].
  Available: \url{https://arxiv.org/pdf/2009.06775.pdf}
\BIBentrySTDinterwordspacing

\bibitem{luong2015effective}
M.-T. Luong, H.~Pham, and C.~D. Manning, ``Effective approaches to
  attention-based neural machine translation,'' 2015.

\bibitem{Ren2019}
Z.~{Ren}, Z.~{Chen}, and S.~{Xu}, ``Triplet based embedding distance and
  similarity learning for text-independent speaker verification,'' in
  \emph{2019 Asia-Pacific Signal and Information Processing Association Annual
  Summit and Conference (APSIPA ASC)}, 2019, pp. 558--562.

\bibitem{rezende2016variational}
D.~J. Rezende and S.~Mohamed, ``Variational inference with normalizing flows,''
  2016.

\bibitem{Schuster1997}
M.~{Schuster} and K.~K. {Paliwal}, ``Bidirectional recurrent neural networks,''
  \emph{IEEE Transactions on Signal Processing}, vol.~45, no.~11, pp.
  2673--2681, 1997.

\bibitem{graves2005}
A.~Graves, S.~Fern{\'a}ndez, and J.~Schmidhuber, ``Bidirectional lstm networks
  for improved phoneme classification and recognition,'' in \emph{Artificial
  Neural Networks: Formal Models and Their Applications -- ICANN 2005}.\hskip
  1em plus 0.5em minus 0.4em\relax Springer Berlin Heidelberg, 2005, pp.
  799--804.

\bibitem{Kingma2019}
D.~P. {Kingma} and M.~{Welling}, \emph{An Introduction to Variational
  Autoencoders}, 2019.

\bibitem{lorenzotrueba2019}
J.~Lorenzo-Trueba, T.~Drugman, J.~Latorre, T.~Merritt, B.~Putrycz,
  R.~Barra-Chicote, A.~Moinet, and V.~Aggarwal, ``Towards achieving robust
  universal neural vocoding,'' 2019.

\bibitem{Uday2019}
\BIBentryALTinterwordspacing
U.~Kamath, J.~Liu, and J.~Whitaker, \emph{Deep Learning for {NLP} and Speech
  Recognition}.\hskip 1em plus 0.5em minus 0.4em\relax Springer, 2019.
  [Online]. Available: \url{https://doi.org/10.1007/978-3-030-14596-5}
\BIBentrySTDinterwordspacing

\bibitem{sptk}
S.~Imai, T.~Kobayashi, and K.~Tokuda, ``Speech signal processing toolkit
  ({SPTK}) - version 3.11,'' \url{https://sp-tk.sourceforge.net}, 2017.

\bibitem{SOX}
``Sound e{X}change ({SOX}),'' \url{https://sox.sourceforge.net}, 2015.

\end{thebibliography}

\end{document}